\documentclass[amsmath,amssymb, aps, prl, twocolumn, notitlepage]{revtex4-1}

\usepackage{graphicx}
\usepackage{dcolumn}
\usepackage{bm}
\usepackage[usenames]{color}
\usepackage{slashed}
\usepackage[utf8]{inputenc}
\usepackage{amsfonts}
\usepackage{amsmath}
\usepackage{amssymb}
\usepackage{hyperref}
\usepackage{latexsym}
\usepackage{color}

\begin{document}
\title{Emergent channel over a pair of pockets in strong density waves}
\author{Di-Zhao Zhu}
\affiliation{International Center for Quantum Materials, School of Physics, Peking University, Beijing, 100871, China}
\author{Yi Zhang}
\email{frankzhangyi@gmail.com}
\affiliation{International Center for Quantum Materials, School of Physics, Peking University, Beijing, 100871, China}

\begin{abstract}
Different channels over which electrons scatter between parts of the Fermi surface are the key to various electronic quantum matters, such as superconductivity and density waves. We consider an effective model in higher dimensions where each of the two pockets in the original model maps to (the Landau levels of) two Dirac fermions. We discover an emergent channel when two Dirac fermions from different pairs annihilate, where the presence of a strong density wave is essential. We support our analysis with numerical calculations on model examples in the vicinity of ferromagnetic and antiferromagnetic orders. We also discuss interesting consequences on electron interaction channels that  beyond-mean-field fluctuations may induce.
\end{abstract}

\maketitle

\emph{Introduction--} Superconductors are extraordinary electronic quantum matters where the electrical resistance vanishes, and magnetic flux fields are expelled from the materials. Microscopically, the electrons near the Fermi surfaces form Cooper pairs following an electron pairing channel \cite{BCS1957, Shankar1994}, where a pair of electrons scatter elastically into another pair with conserved momentum, e.g., via attractive electron interactions \cite{Scalettar1989} or electron-phonon coupling \cite{BCS1957, Shankar1994}. The discoveries of high-temperature superconductors in Cu-based \cite{Bednorz1986} and Fe-based \cite{Takahashi2008, Dagotto2013} materials bring hope for room-temperature superconductors. The unlikelihood of an electron-phonon mechanism calls for unconventional pairing mechanism \cite{Scalapino1492}, which remains controversial after more than three decades of research and hindering the search for enhanced $T_c$.

The situation becomes even more complicated when various intertwined orders are established in the phase diagrams in the vicinity of high-temperature superconductivity \cite{KivelsonIntertwine, Efetov2013}, including ferromagnetic order \cite{Xu2009}, antiferromagnet order \cite{Inosov2010, Dai2012, Dai2015, LuChen2015}, charge density waves \cite{Wu2011, zhang2018using, Badoux2016, Neto2015, Chang2012}, pair density waves \cite{JCDavis2016PDW}, nematic order \cite{Chu2010, Wang2015}, etc. Interestingly, electron scattering channels across a nesting wave vector may also be responsible for density waves \cite{Peierls, Lieb1987}. However, the theory is still incomplete as a nested Fermi surface cannot explain the origin of density waves in various scenarios \cite{Frankicdw2015, Zhu2015}.

\begin{figure}
\includegraphics[width=0.98\linewidth]{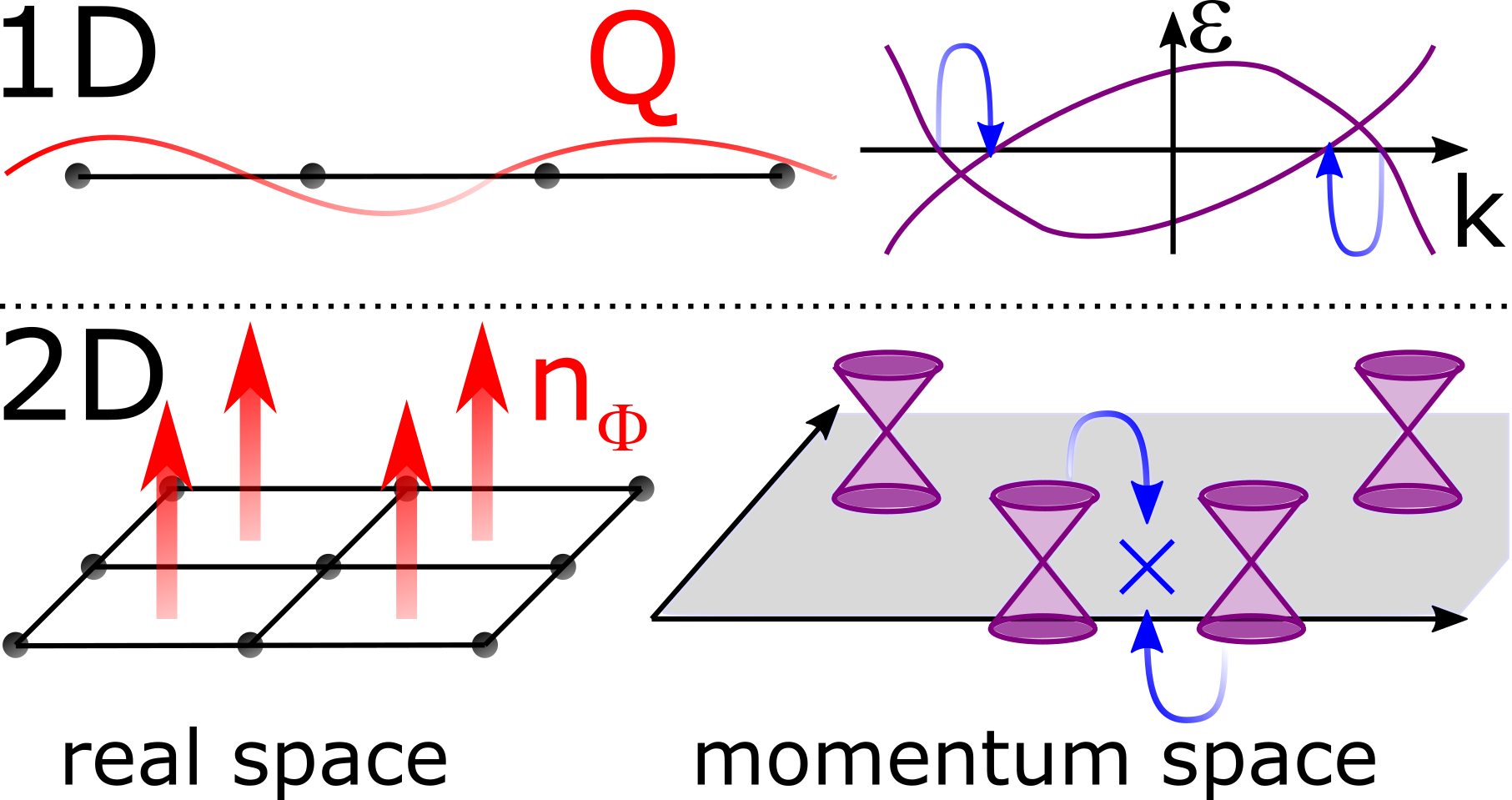}
\caption{(1) A 1D system with a density wave is equivalent to an effective 2D system with a magnetic field. (2) In the momentum space, each of the 1D system's two pockets maps to (the Landau levels of) a pair of Dirac fermions in the effective 2D system. (3) A Dirac fermion may annihilate with another from a different pair when the density wave strength exceeds a threshold. (4) The two original electron scattering channels across their respective pocket's nesting wave vector merge into a single, exotic channel.}
\label{fig:scheme}
\end{figure}

Here, we analyze the characteristics of electronic quantum matters in both weak and strong density waves in a unified theory. Our theory scheme relies on an effective model in higher dimensions where each of the original model's two pockets maps to (the Landau levels of) two Dirac fermions, where the small parameter is the density wave vector $Q$ or $Q-\pi$ instead of the density wave strength. While the results in weak density waves are consistent with the perturbative Fermi-surface-nesting picture, we discover an emergent channel that engages both pockets simultaneously in strong density waves, where two Dirac fermions from different pairs annihilate as we illustrate in Fig. \ref{fig:scheme}. Applications of our theory to density-wave systems near ferromagnetic and antiferromagnetic orders show full consistency with numerical calculations. Further, this emergent channel may evolve into interesting electron interaction channels once we start to take fluctuations of the density wave strength beyond the mean-field theory into consideration.

\begin{figure}
\includegraphics[width=.98\linewidth]{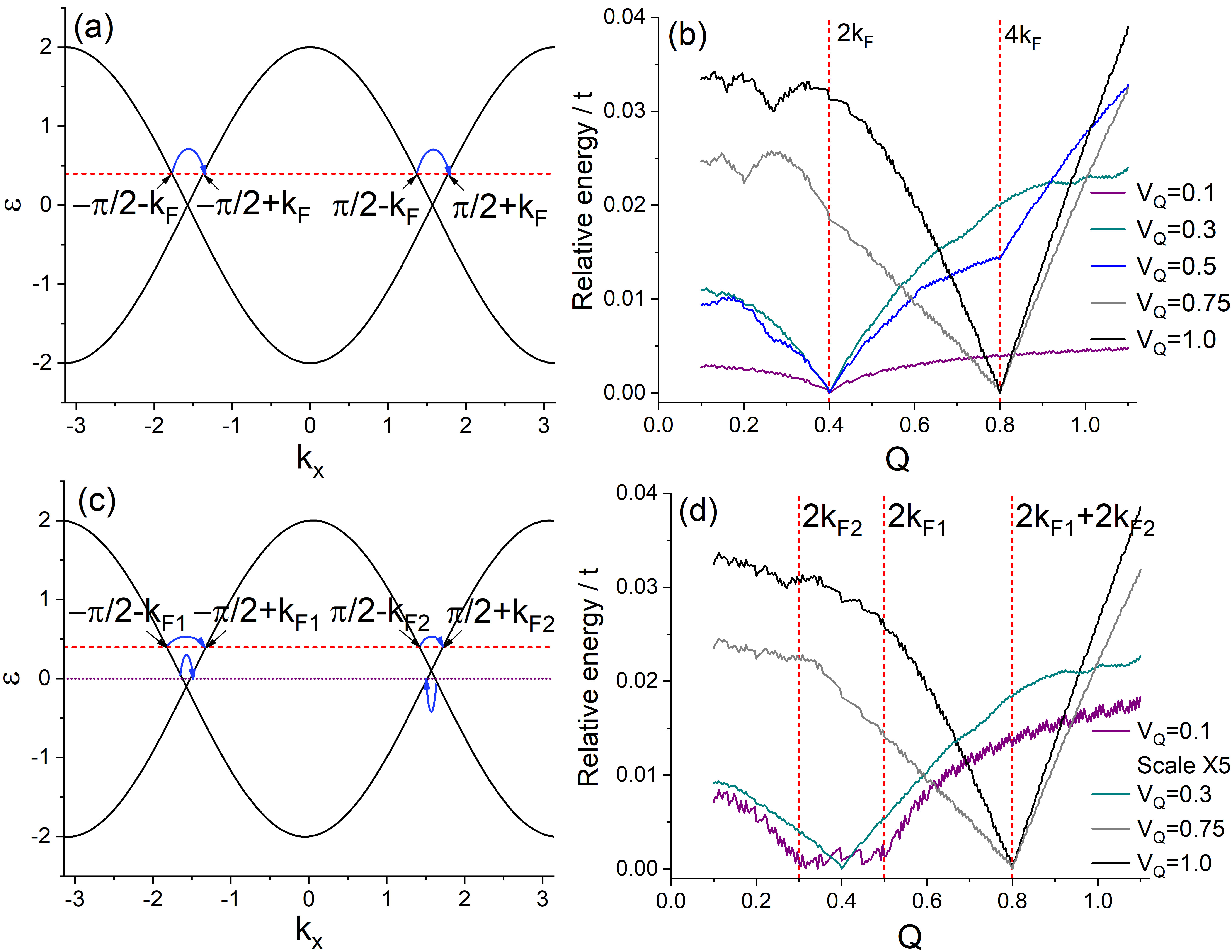}
\caption{(a) The dispersion of $H_0$ with $t'=0$.
The red dashed line shows the Fermi level where $k_{F}=0.2$,
and the blue arrows denote the electron scattering channels across the $2k_{F}$ nesting vector.
(b) The average electron energy of $H_{1D}$ with $H_0$ in (a) versus the wave vector $Q$. Raising $V_Q$ to $O(1)$, the optimal $Q_{opt}$ switches to an anomalous $4k_F$. (c) The dispersion with a small $t'=0.05$ and same $n$ as (a) gives two pockets with $k_{F1,2}=0.2\pm 0.05$.
(d) The average electron energy of $H_{1D}$ with $H_0$ in (c) suggests the optimal $Q_{opt}=2k_{F1}+2k_{F2}$ at large $V_Q$.
}
\label{fig:H0disp}
\end{figure}

\emph{Model example near ferromagnetic order--} Without loss of generality, we consider the following one-dimensional (1D) tight-binding model $H_{1D}=H_{0}+H_{DW}$:
\begin{eqnarray}
H_{0}&=&\sum_{x} \left(it'-t\right) c_{x+1,\uparrow}^{\dagger}c_{x,\uparrow}+\left(it'+ t\right)c_{x+1,\downarrow}^{\dagger}c_{x,\downarrow}+\mbox{h.c.}, \nonumber \\
H_{DW}&=&\sum_{x,s,s'}\left[\epsilon_0 -2V_Q\cos \left(Qx+\varphi_0\right)\right] c_{x,s}^{\dagger} \sigma^z_{ss'}c_{x,s'} \nonumber \\
& &+ 2\lambda_Q \sin \left(Qx+\varphi_0\right)
c_{x,s}^{\dagger} \sigma^x_{ss'}c_{x,s'}, \label{eq:H1d}
\end{eqnarray}
where $c_{x,\uparrow}$ and $c_{x,\downarrow}$ are the annihilation operators for spin-up and spin-down electrons, respectively. $H_0$ is a nearest-neighbor-hopping lattice model explicitly invariant under lattice translations, which are subsequently broken by the density waves in $H_{DW}$. The system has a fixed electron density $n=n_\uparrow+n_\downarrow=1+(2k_{F1}+2k_{F2})/2\pi$, where $\left|k_{Fi}\right|\ll O(1)$ is the Fermi vector of the $i^{th}$ pocket ($k_{Fi}>0$ ($k_{Fi}<0$) for an electron (hole) pocket) given by the $H_0$ dispersion, see Figs. \ref{fig:H0disp}a and \ref{fig:H0disp}c. For convenience, we set the typical hoping amplitude $t=1$ as our unit of energy. Also, we can include a small $t'$ to break the particle-hole symmetry so that the two pockets may differ from each other with distinct nesting wave vectors $2k_{F1}$ versus $2k_{F2}$.  $\epsilon_0$ is the strength of a long-range ferromagnetic order along $\sigma_z$. Near critical point, the $Q=0$ ferromagnetic order may become relatively shorter-ranged, and spin-wave fluctuations develop and soften near $Q\sim 0$. In a similar spirit to the Peierls transition, we take a mean-field stance and treat the strengths  $2V_Q$ ($2\lambda_Q$) of the $\sigma^z$ ($\sigma^x$) component of the spin density waves as constants, and search for the wave vector $Q_{opt} \ll2\pi$ that yields the lowest systematic energy. We note that a minimal Dirac-fermion model necessitates the indices $s=\uparrow,\downarrow$, which can be interpreted as spins, pseudo-spins, orbitals, etc., so that $H_{DW}$ describes different density wave scenarios. In the absence of the degree of freedom $s$, however, the system reduces to conventional Landau levels. We will discuss the differences as well as fluctuation effects later.

Numerical results on the average electron energy of $H_{1D}$ as a function of $Q$ are summarized in Fig. \ref{fig:H0disp}.
Fig. \ref{fig:H0disp}a (\ref{fig:H0disp}c) plots the dispersion of $H_0$ with $t'=0$ ($0.05$), which has two crossings at $k=\pm \pi$. The red dashed lines show Fermi levels for the same electron density $n=1+0.8/2\pi$, and the blue arrows denote the electron scattering channels across the $2k_{Fi}$ nesting vectors. Besides, the purple dotted line shows the Fermi level for electron density $n\sim 1$ close to charge neutrality, where $k_{F2}<0$ for the hole pocket in our convention. Figs. \ref{fig:H0disp}b and \ref{fig:H0disp}d exhibit the average electron energy of $H_{1D}$ versus the wave vector $Q$ for several different density wave strengths $V_Q$. We set $\epsilon_0=2V_Q$ and $\lambda_Q=V_Q$ in our numerical calculations on chains $L\sim 4000$ with well-controlled boundary conditions. Energies are plotted relative to their minimums for clarity and scaled up if necessary.
For weak density waves, the energy is minimum at the nesting values of $Q_{opt}=2k_{Fi}$, $i=1,2$, consistent with the Fermi surface nesting and $H_{DW}$ as a perturbation. The resulting electronic quantum matter is dominated by the electron scattering channels $c^\dagger_{\pi/2+k_{F2},\uparrow} c_{\pi/2-k_{F2},\downarrow}$ and $c^\dagger_{-\pi/2+k_{F1},\downarrow} c_{-\pi/2-k_{F1},\uparrow}$ with momentum transfers $2k_{F2}$ and $2k_{F1}$, respectively. In strong density waves, on the contrary, the energy minimum clearly shows a single optimal wave vector $Q_{opt}=2k_{F1}+2k_{F2}$, while the features at the original nesting values $2k_{F1}$ and $2k_{F2}$ are largely suppressed, indicating that the separate scattering channels across their respective nesting vectors are no longer available, and merge into a single, new channel with momentum transfer $2k_{F1}+2k_{F2}$. The overall low-energy electronic degrees of freedom are reduced. What is the mechanism of this emergent channel with anomalous $Q_{opt}$ and the property of the resulting electronic quantum matter?

\emph{Theory--} The emergent channel occurs at relatively large density wave strengths. Therefore, we take the following theoretical approach for controlled analysis. For incommensurate $Q$, the physics of Eq. \ref{eq:H1d} is independent of $\varphi_0$. If we regard $\varphi_0\equiv k_y$ as the momentum in an extra $\hat y$ dimension and sum over $k_y$, the resulting two-dimensional (2D) system with a magnetic field of flux density $n_\Phi = Q/2\pi$ is equivalent to those of the original $H_{1D}$ \cite{Frankicdw2015}. Without $n_\Phi$, the 2D effective model is translation invariant and diagonalizable in the $\vec{k}=(k_x, k_y)$ basis $\sum_{\vec{k}}\left(c^\dagger_{\vec{k}\uparrow}, c^\dagger_{\vec{k}\downarrow} \right) h_{2D}\left(\vec{k}\right) \left(c_{\vec{k}\uparrow}, c_{\vec{k}\downarrow}\right)^T$:
\begin{eqnarray}
h_{2D}(\vec{k})&=&\sigma^{z}\left(\epsilon_0-2V_Q\cos k_{y}-2t\cos k_{x}\right)+\sigma^{x}2\lambda_Q\sin k_{y} \nonumber \\ & & + 2t' \sin k_x.
\label{eq:H2dk}
\end{eqnarray}

\begin{figure}
\includegraphics[width=1.0\linewidth]{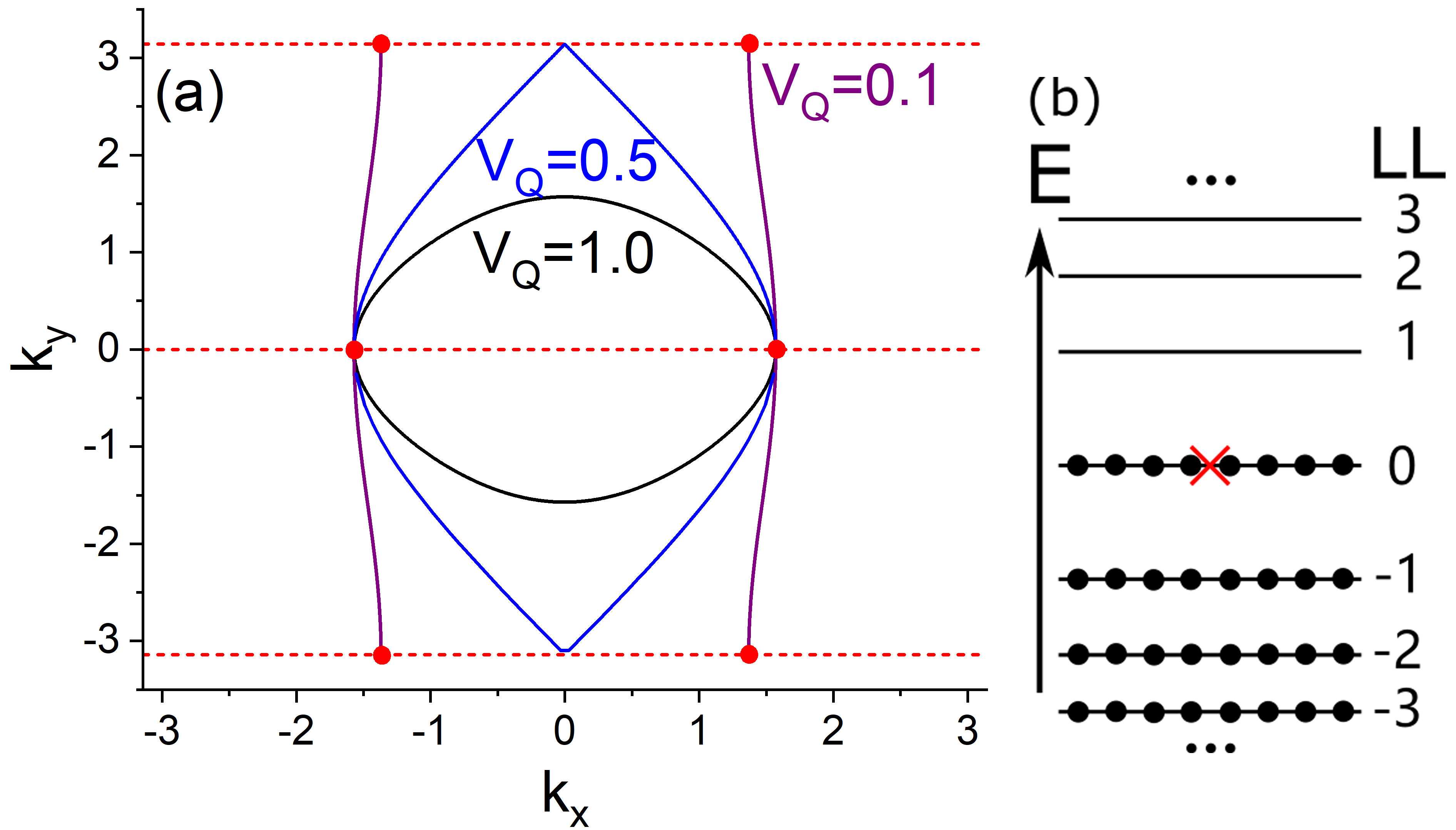}
\caption{(a) The red dots show the locations of the Dirac fermions in the Brillouin zone of the 2D effective model in Eq. \ref{eq:H2dk}. At $V_Q=0.5$, the contour of $\epsilon_0-2V_Q \cos k_y - 2\cos k_x = 0$ reduces its crossings with $k_y = 0, \pm \pi$ (the red dashed lines) from $n_D=4$ in weak density waves to $n_D=2$ in strong density waves. $\epsilon_0 = 2V_Q$. (b) The Landau levels of a Dirac fermion center around the zeroth Landau level. Above charge neutrality (denoted by the red cross), the optimal magnetic field is to fill the zeroth Landau level and leave all the above Landau levels empty.}
\label{fig:EvsQ}
\end{figure}

\begin{figure}
\includegraphics[width=1.\linewidth]{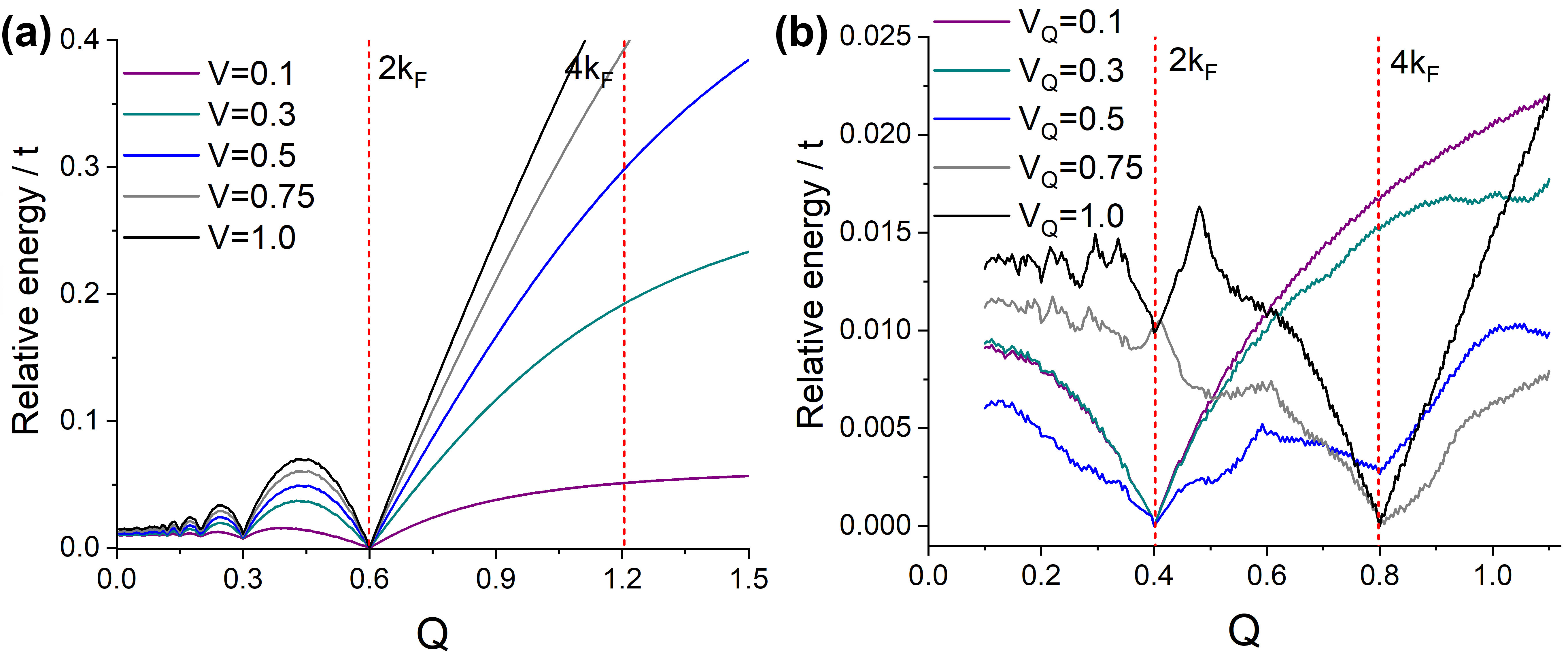}
\caption{(a) The average electron energy versus the density wave $Q$ for Eq. \ref{eq:convQ} shows that the minimum energy is always at $Q_{opt}=2k_F$ when conventional Landau level physics is at play. Here, we set $\epsilon_k=-2\cos k_x$, $k_F=0.3$, and $L\sim 4000$. (b) Similar to Fig. \ref{fig:H0disp}b despite a relatively small $\lambda_Q=0.25$, the average electron energy of $H_{1D}$ in Eq. \ref{eq:H1d} versus the wave vector $Q$ also shows that the optimal $Q_{opt}$ with minimum energy changes from the nesting value $Q_{opt}=2k_F$ to an anomalous $Q_{opt}=4k_F$.}
\label{fig:EvsQconven}
\end{figure}

For the model parameters used in Fig. \ref{fig:H0disp}, this two-band model $h_{2D}$ has $n_D=4$ ($n_D=2$) Dirac fermions when $V_Q$ is small (large), where two Dirac fermions at $k_y=\pm\pi$ annihilate when $V_Q=0.5$, see Fig. \ref{fig:EvsQ}a. $n_D$ is the number of Dirac fermions irrespective of the details. In a magnetic field, a 2D Dirac fermion forms Landau levels $\epsilon_N \propto \pm \sqrt{N}$ centered at the Dirac node, see Fig. \ref{fig:EvsQ}b. We deem the Dirac fermions as independent and neglect the quantum tunneling (magnetic breakdown \cite{Blount1960, Cohen1961, Frankicdw2015}) between them, which is a controlled approximation when the Dirac nodes are far apart, and $Q\ll 2\pi$ is small. Consequently, the optimal magnetic field $\propto Q_{opt}$ at electron density $n=1+4k_F/2\pi$ is to fill all the zeroth Landau levels and leave all the higher Landau levels empty. For minimum energy, higher Landau levels are evacuated to avoid finite-energy excitations $\propto \sqrt{Q}$, and the zeroth Landau levels are fully filled instead of superfluously vacant at the expense of a larger magnetic field $\propto Q$ that lifts the energy of the Fermi sea. Such periodicity of the systematic energy versus Landau level fillings constitute the premise of quantum oscillations, e.g., the dHvA effect \cite{Onsager1952,Lifshitz1956}. As a result, the electron density above charge neutrality should match half of the zeroth Landau levels' degeneracy from all Dirac fermions:
\begin{equation}
\frac{4k_F}{2\pi}=n-1=\frac{n_D}{2}\frac{Q_{opt}}{2\pi}.
\label{eq:LLfill}
\end{equation}
Thus, we have $Q_{opt}=2k_F$ in weak density waves consistent with the Fermi surface nesting theory and an anomalous $Q_{opt}=4k_F$ in strong density waves. Also, as long as the shift to the $h_{2D}(\vec k)$ dispersion, e.g., the $2t'\sin(k_x)$ term, is not large enough to shuffle the zeroth and non-zero Landau levels, the condition for optimal filling in Eq. \ref{eq:LLfill} holds, and we have $Q_{opt}=2k_{F1}+2k_{F2}$ for $n_D=2$. These conclusions are fully consistent with our numerical results. We emphasize the Dirac fermions' crucial role in the emergent channel with anomalous $Q_{opt}$, which is absent in conventional quadratic dispersions. For instance, we can map
\begin{equation}
H_{1D} = \sum_k \epsilon_k c^\dagger_k c_k + \sum_x V\cos\left(Qx+k_y\right)
\label{eq:convQ}
\end{equation}
to an effective single-band 2D model with electron density $n= 2k_F/2\pi \ll 1$ and a magnetic field of flux density $n_\Phi=Q/2\pi$. Generally, the most energy-favorable $Q_{opt}$ simply fills the lowest Landau level and nothing else:
\begin{equation}
\frac{n}{n_\Phi}=\frac{2k_F}{Q_{opt}}=1.
\end{equation}
Therefore, the usual nesting condition $Q_{opt}=2k_F$ holds universally irrespective of weak or strong density waves \footnote{Research on scenarios when the local minimums for filling the $N=2, 3, \cdots$ lowest Landau levels become dominant is currently in progress.}. Indeed, numerical results for the nearest-neighbor tight-binding model $\epsilon_k=-2\cos k_x$ suggest that the energy is minimal at $Q_{opt}=2k_F$ across the probed range of $V$, see Fig. \ref{fig:EvsQconven}a.

\emph{Properties of the electronic quantum matter with the emergent channel--} (1) $V_Q$ needs to be comparable with the typical hopping amplitude $t=1$ and large (due to large density wave order parameters and/or strong coupling), a parameter region typically unattainable via perturbative approaches. (2) $\lambda_Q$ needs to be nonzero so that the Dirac points at $k_y=0, \pi$ remain separate in the presence of a magnetic field $\propto Q$ by a barrier $2\lambda_Q\sin k_y$. Since the magnetic breakdown between the Dirac points occurs at around $Q\sim O(v_y/v_x)\sim O(\lambda_Q)$ and above \cite{Patrick2017}, $\lambda_Q$ does not need to be large, yet its presence is essential, see Fig. \ref{fig:EvsQconven}b for example. (3) The resulting system is fully incompressible with a finite excitation gap (the Landau level spacing in the effective model) $\propto \sqrt{Q_{opt}} \gg O(Q_{opt})$ for small $k_F$, and relatively stable towards perturbations such as thermal fluctuations $k_B T\ll \sqrt{Q_{opt}}$. (4) The remaining pair of Dirac fermions in the effective theory descend from two separate pockets in the original model in strong density waves, suggesting that the entire range of $k_x$ in between is physically involved and a Fermi-liquid-theory consideration may no longer be valid. In contrast, each pair of Dirac fermions are still around the $k_x$ of the original 1D pockets in weak density waves, limiting the low-energy physics to these $k_x$.

\begin{figure}
\includegraphics[width=0.98\linewidth]{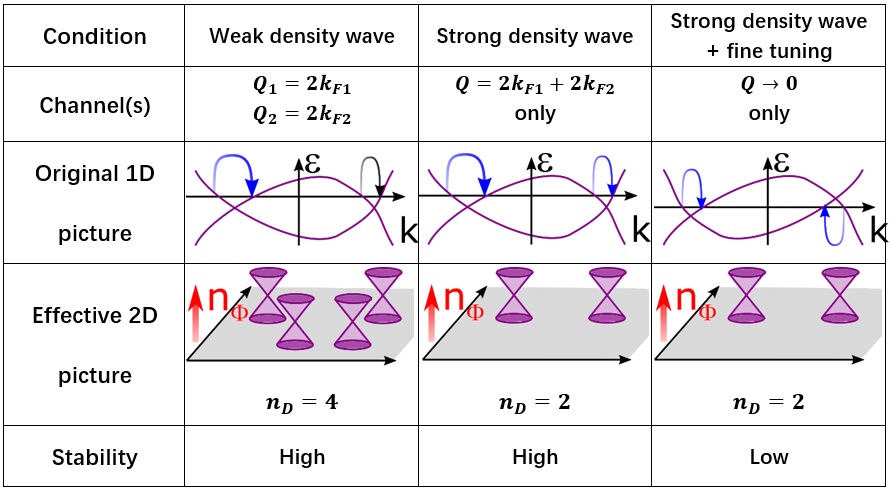}
\caption{Each 1D pocket corresponds to a pair of Dirac fermions in the 2D theory and contributes a separate electron scattering channel in weak density waves. In strong density waves, two fermions from different pairs annihilate, signaling the merge of the previous two channels into an emergent channel with momentum transfer $Q=2k_{F1}+2k_{F2}$. The emergent channel may conserve momentum with fine-tuning, as the electron pocket and the hole pocket in the 1D dispersion show opposite nesting vectors.}
\label{fig:phasediagram}
\end{figure}

An interesting special case is when we have a hole pocket and an electron pocket with similar sizes in the initial $H_0$, e.g., the electron density corresponds to the purple dotted line in Fig. \ref{fig:H0disp}c. Then, the anomalous $Q_{opt}=2k_{F1}-\left|2k_{F2}\right|\rightarrow 0$ and the momentum transfer of the emergent channel vanishes. For our argument in Eq. \ref{eq:LLfill} to be valid through the $Q\rightarrow 0$ limit \footnote{We have applied the Landau level filling argument before the $Q\rightarrow 0$ limit, similar to $\omega \rightarrow 0$ before $k \rightarrow 0$ order of limits for the superfluid density \cite{Scalapino1992, Scalapino1993}.}, the two remaining Dirac nodes and thus their zeroth Landau levels should be at the same energy \footnote{Effectively, the carrier density difference between the original electron pocket and hole pocket is attributed to the two annihilated Dirac fermions. Note that it is conceivable that the emergent channel survives, to some extent, beyond these conditions, which is only for theoretical rigorousness.}. Later, we will show such an example in Eq. \ref{eq:H1dAFM}. Since the Landau level spacings $\propto \sqrt{Q_{opt}}$ become small as $Q_{opt}\rightarrow 0$, the emergent channel is not as stable under this special condition. We summarize different scenarios, the physical pictures, and related properties in Fig. \ref{fig:phasediagram}.

Expanding the Hamiltonian in Eq. \ref{eq:H2dk} to linear orders $\vec q = \vec k - \vec k_0$ around each Dirac node $\vec k_0$, we have $h_{2D}(\vec q) = \sigma^z v_x q_x + \sigma^x v_y q_y$, where $v_x$ and $v_y$ are the respective Fermi velocities (linear-$k$ differentials) and their signs determine the chirality of the corresponding Dirac fermion \cite{Goerbig2017}. In the effective magnetic field, the commutation relation $[q_x, q_y]=[k_x, k_y]=[k_x, Qx]=-iQ$ defines the ladder operators $\hat a = \sqrt{|v_x|/2Q|v_y|} q_x - i \sqrt{|v_y|/2Q|v_x|} q_y$ with $[\hat a, \hat a^\dagger]=1$, which we may use to re-express
\begin{equation}
h_{2D}(\vec q) = \sqrt{2Q|v_x v_y|} [\frac{\sigma^z \mbox{sgn}(v_x) + i\sigma^x \mbox{sgn}(v_y)}{2}\hat a+\mbox{h.c.}]
\end{equation}
for the Dirac-fermion Landau levels. Notably, $(\sigma^z \pm i \sigma^x)/2$ is the raising and lowering operators among the $\sigma^y$ states, thus there exists a zeroth Landau level fully polarized in either $\sigma^y$ or $-\sigma^y$ depending on the chirality $\mbox{sgn}(v_x) \mbox{sgn}(v_y)$ yet irrespective of the details ($k_0$, $|v_x|$, and $|v_y|$) of the Dirac fermion.

Since the low-energy degrees of freedom and the subsequent emergent density waves are associated with the zeroth Landau level, they are limited to the corresponding $\sigma^y=\pm 1$ sectors, labeled as $s=\rightarrow,\leftarrow$, respectively. As the two Dirac fermions in the center annihilate and corresponding degrees of freedom move to higher energy in strong density waves, the remaining low-energy degrees of freedom and the momentum transfer of $2k_{F1}+2k_{F2}$ imply that the emergent channel should relate to $c^\dagger_{\pi/2+k_{F2},\rightarrow}c_{\pi/2-k_{F2},\rightarrow}c^\dagger_{-\pi/2+k_{F1},\leftarrow}c_{-\pi/2-k_{F1},\leftarrow}$. At the mean-field level, such a four-fermion term fully reduces to its fermion-bilinear descendants following Wick's theorem. Beyond the mean-field theory, e.g., in the presence of $V_Q$ fluctuations, the four-fermion interaction may achieve independent implications and describe scattering an electron pair into another.

\emph{Model example near antiferromagnetic order--} For density waves with wave vectors too large for a Landau-level point of view, we may separate $Q$ into a large, commensurate component and a small, incommensurate component $q\ll 2\pi$. Take the spin density waves near antiferromagnetic order $Q=\pi+q$, $q/2\pi \ll O(1)$ as an example:
\begin{eqnarray}
H'_{0}&=&\sum_{x,s} - t c_{x+1,s}^{\dagger}c_{x,s}- it' c_{x+3,-s}^{\dagger}c_{x,s} +\mbox{h.c.} \nonumber\\
H'_{DW}&=&\sum_{x,s,s'}(-1)^x\left[\epsilon_0 -2V_q\cos \left(qx+\varphi_0\right)\right] c_{x,s}^{\dagger}\sigma^z_{ss'}c_{x,s'} \nonumber \\
& &+ \left[\epsilon'_0-2\lambda_q \sin \left(qx+\varphi_0\right)\right] c_{x,s}^{\dagger}\sigma^x_{ss'}c_{x,s'}.
\label{eq:H1dAFM}
\end{eqnarray}
Then, we define $a_{x,s}=c_{x,s}$ for even $x$ and $a_{x,s}=c_{x,-s}$ for odd $x$, $s=\uparrow,\downarrow$, which removes the $(-1)^x$ factor in $H'_{DW}$, and $H'_0$ becomes:
\begin{eqnarray}
H'_{0}=\sum_{x,s,s'} - t a_{x+1,s}^{\dagger}\sigma^x_{ss'} a_{x,s}
- it' a^\dagger_{x+3,s} \sigma^0_{ss'} a_{x,s'} +\mbox{h.c.}. \nonumber \\ \label{eq:H1dAFM2}
\end{eqnarray}
Like before, we define $\varphi_0 \equiv k_y$ and map $H'_{0}+H'_{DW}$ to a 2D effective model:
\begin{eqnarray}
h'_{2D}(\vec{k})&=& \sigma^{x}\left(\epsilon'_0-2\lambda_q \sin k_{y} -2t\cos k_{x}\right) \nonumber \\
& &+\sigma^{z}\left(\epsilon_0-2V_q \cos k_{y}\right)- 2t'\sin 3k_x,
\label{eq:H2dkAFM}
\end{eqnarray}
with a magnetic field of flux density $n_\Phi=q/2\pi \ll O(1)$.

\begin{figure}
\includegraphics[width=.98\linewidth]{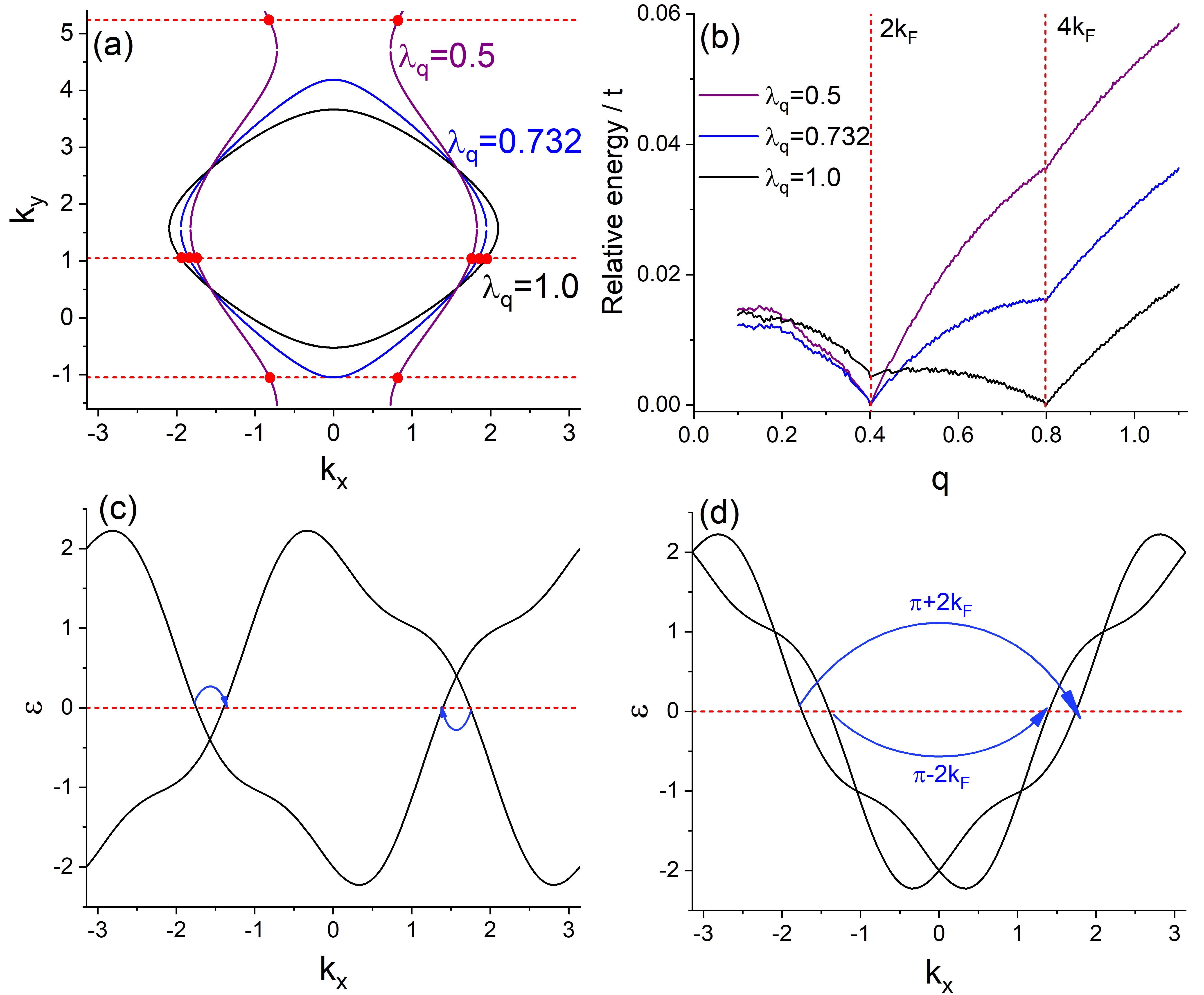}
\caption{(a) The red dots show the locations of the Dirac fermions of the model in Eq. \ref{eq:H2dkAFM}. At $\lambda_q=0.732$, the contour of zero $\sigma^x$ coefficient 
reduces its crossings with $k_y = \pm \pi/3$ (the red dashed lines) from $n_D=4$ to $n_D=2$.
(b) The average electron energy of the model in (a) versus the wave vector $q=Q-\pi$ for various $\lambda_q$ shows that the optimal $q_{opt}$ changes from the nesting value of $q_{opt}=2k_F$ to an anomalous value of $q_{opt}=4k_F$.
(c) The electron scattering channels over the electron pocket and the hole pocket have opposite momentum transfers. (d) Same as (c) but in the $c$-operator representation.}
\label{fig:afmcase}
\end{figure}

This two-band model $h'_{2D}$ has $n_D=4$ Dirac nodes when $\lambda_q$ is small and $n_D=2$ Dirac nodes when $\lambda_q$ is large, see Fig. \ref{fig:afmcase}a. For illustration, we set $\epsilon_0=V_q$, $\epsilon'_0=\lambda_q$, $n=1+4k_F/2\pi$, $k_F=0.2$ and $t'=0$ in our numerical calculations on chains $L\sim 4000$ chains with well-controlled boundary conditions. At density wave strength $\lambda_q=\sqrt{3}-1$ comparable with hopping parameters, the two Dirac fermions at $k_y=-\pi/3$ annihilate, and the emergent channel with anomalous $q_{opt}=4k_F$ takes over the original nesting value $q_{opt}=2k_F$ following Eq. \ref{eq:LLfill}. Our numerical results summarized in Fig. \ref{fig:afmcase}b show full consistency with our theoretical analysis.

\begin{figure}
\includegraphics[width=1.\linewidth]{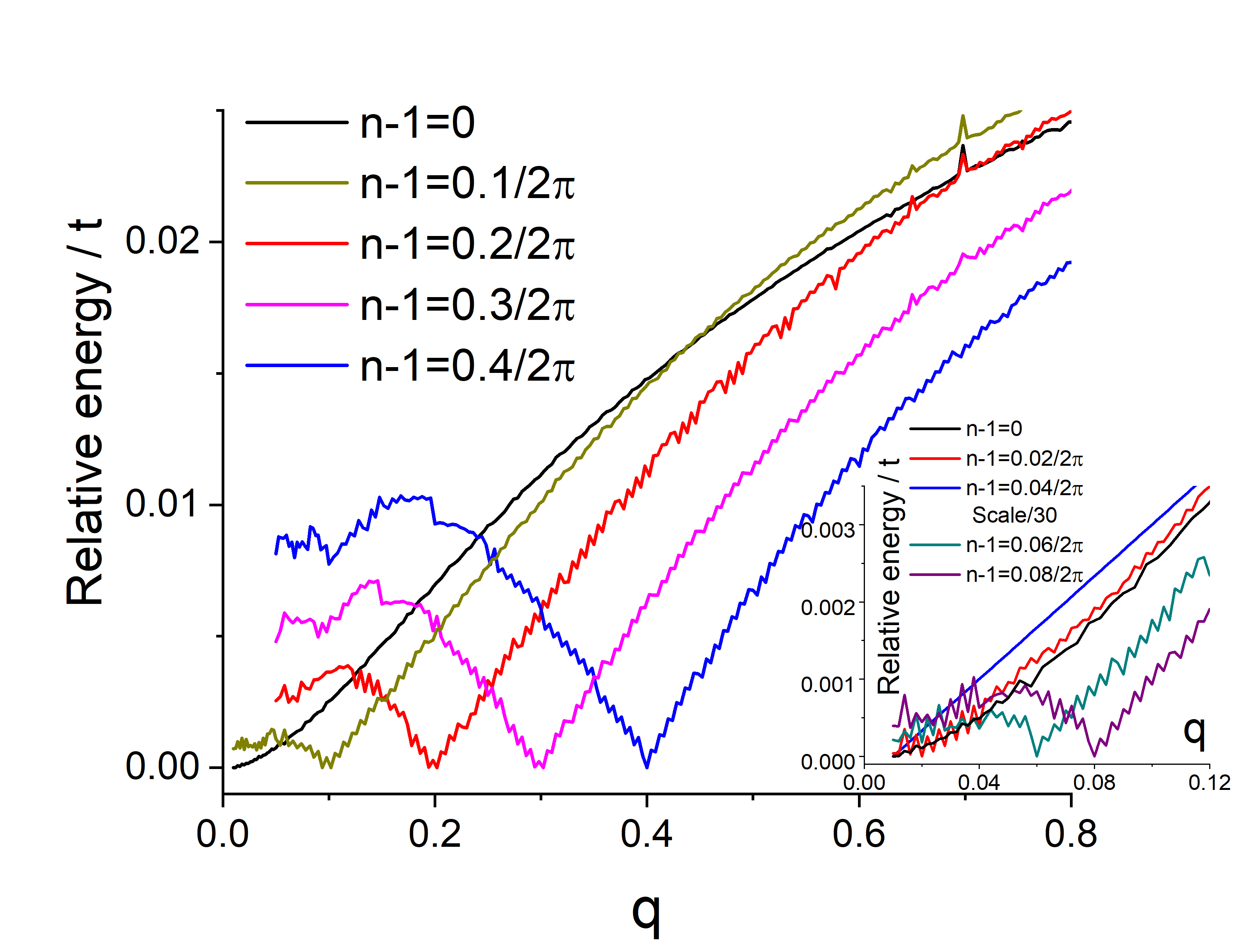}
\caption{The average electron energy versus the wave vector $q$ shows the momentum transfer of the emergent channel $q=2k_{F1}-\left|2k_{F2}\right|=2\pi\left(n-1\right)$ approaches 0 as the electron density $n \rightarrow 1$ in a strong density wave $V_q=1.0$, $\lambda_q=1.366$. The original 1D dispersion has an electron pocket and a hole pocket, see Fig. \ref{fig:afmcase}c. Inset: zoom-in data at smaller $q$ obtained on longer chains.}
\label{fig:qgoto0}
\end{figure}

Next, we show an example where the emergent channel conserves momentum. Here, we exploit the $k_x$ displacement of the 2D theory's Dirac nodes away from the original 1D pockets in strong density waves: the two Dirac nodes are at $k_x=\pm 2\pi/3$ and $k_y=\pi/3$ for $V_q=1$ and $\lambda_q=(\sqrt{3}+1)/2$; the term $2t'\sin 3k_x$, $t'=0.2$ does not affect the Dirac nodes, yet gives rise to an asymmetric shift and thus two pockets with opposite carrier types in the 1D dispersion when the electron density $n \sim 1$, see Fig. \ref{fig:afmcase}c. In strong density waves, the two separate scattering channels with opposite momentum transfers $2k_{F1}$ and $2k_{F2}\sim -2k_{F1}$ merge into an emergent channel with vanishing momentum transfer $q_{opt}=2k_{F1}-\left|2k_{F2}\right|\rightarrow 0$. Numerical results summarized in Fig. \ref{fig:qgoto0} confirm our analysis.

At last, we perform the inverse transformation back to the original $c$-fermion representation (with $k_x \rightarrow k_x+\pi$ transformation for $\sigma^x=-1$), which shows two much larger pockets, see Fig. \ref{fig:afmcase}d. Therefore, two small pockets (with opposite carrier types) are not fully necessary for the emergent channel (with momentum conservation). However, the emergent channel only acts on those Fermi points approximately connected by half reciprocal lattice wave vector in the $c$-fermion representation, thus keeping the strict condition on the electron density or the Fermi energy. For instance, we may consider another example similar to the first $H_0$ in Eq. \ref{eq:H1d}:
\begin{eqnarray}
H''_{0}&=&\sum_{x}- t c_{x+1,\uparrow}^{\dagger}c_{x,\uparrow}+ t c_{x+1,\downarrow}^{\dagger}c_{x,\downarrow}+\mbox{h.c.} \label{eq:H1D2ndAFM} \\
& &- t' \left(-1\right)^x \left(c_{x+3,\uparrow}^{\dagger}c_{x,\downarrow}+c_{x+3,\downarrow}^{\dagger}c_{x,\uparrow}   \right) + \mbox{h.c.} \nonumber
\end{eqnarray}
coupled to a similar antiferromagnetic $H'_{DW}$ in Eq. \ref{eq:H1dAFM} with $\sigma_y$ instead of $\sigma_x$, $Q=\pi +q\rightarrow \pi$. After the transformation $a_{x,s}=c_{x,s}$ for even $x$ and $a_{x,s}=ic_{x,-s}$ for odd $x$, $s=\uparrow,\downarrow$, $H''_{0}+H'_{DW}$ transforms into a similar form with Eq. \ref{eq:H2dkAFM} and the discussions afterwards follow through straightforwardly.

\emph{Discussions--} Although we mainly focus our discussions on 1D examples for simplicity, we can generalize our methods and arguments straightforwardly to higher dimensions. The mean-field initial setup would make more sense, and the electronic quantum matter dominated by the emergent channels will be more stable. A heuristic case can decouple along orthogonal directions into multiple 1D systems similar to those in Eq. \ref{eq:H1d} or Eq. \ref{eq:H1dAFM}. Also, our work on two-dimensional density waves from a three-dimensional perspective involving topological nodal lines and Weyl semimetals, etc., is in progress.

Using the wave vectors as our controlled small parameter, our theoretical perspective via Dirac fermion Landau levels offers simple yet elegant insights towards electronic quantum matter properties in strong density waves and the breakdown of conventional perturbative approaches. The signature of the emergent channel is an anomalous $Q_{opt}$ momentum transfer that differs from the nesting vectors and should be sensitive to electron scattering experiments and change of electron density via either doping or gating. Also, additional electron interaction channels open up in the presence of fluctuating strong density waves, offering intuitive alternatives towards strongly-correlated electron quantum matters such as unconventional superconductivity.

\emph{Acknowledgement:} We thank Tian-Lun Zhao, Yan Zhang, Yuan Li, and X.-C. Xie for insightful discussions. YZ is supported by the start-up grant at Peking University. The computation was supported by High-performance Computing Platform of Peking University.

\bibliography{refs}

\end{document}